\documentclass[conference]{IEEEtran}
\IEEEoverridecommandlockouts

\usepackage{cite}
\usepackage{amsmath,amssymb,amsfonts}
\usepackage{algorithmic}
\usepackage{textcomp}
\usepackage{xcolor}
\usepackage{graphicx}
\usepackage{dcolumn}
\usepackage{bm}
\usepackage{braket}
\usepackage{hyperref}
\usepackage{mathtools}
\usepackage{caption}
\usepackage{subcaption}
\usepackage{tabularx}
\usepackage{multirow}
\usepackage{float}
\def\BibTeX{{\rm B\kern-.05em{\sc i\kern-.025em b}\kern-.08em
    T\kern-.1667em\lower.7ex\hbox{E}\kern-.125emX}}
\begin{document}

\title{Few-Shot, Robust Calibration of Single Qubit Gates Using Bayesian Robust Phase Estimation
\thanks{Zhubing Jia is currently with the Department of Physics, The University of Illinois at Urbana-Champaign, Urbana, IL, USA. \\
Corresponding author: Travis Hurant (travis.hurant@duke.edu)}
}

\author{
    \IEEEauthorblockN{
        Travis Hurant\IEEEauthorrefmark{1}\IEEEauthorrefmark{2}, Ke Sun\IEEEauthorrefmark{1}\IEEEauthorrefmark{3}, Zhubing Jia\IEEEauthorrefmark{1}\IEEEauthorrefmark{3}, Jungsang Kim\IEEEauthorrefmark{1}\IEEEauthorrefmark{2}\IEEEauthorrefmark{3}, Kenneth R. Brown\IEEEauthorrefmark{1}\IEEEauthorrefmark{2}\IEEEauthorrefmark{3}\IEEEauthorrefmark{4}
    }
    \IEEEauthorblockA{\IEEEauthorrefmark{1} Duke Quantum Center, Duke University, Durham, NC USA}
    \IEEEauthorblockA{\IEEEauthorrefmark{2} Department of Electrical and Computer Engineering, Duke University, Durham, NC USA}
    \IEEEauthorblockA{\IEEEauthorrefmark{3} Department of Physics, Duke University, Durham, NC USA}
    \IEEEauthorblockA{\IEEEauthorrefmark{4} Department of Chemistry, Duke University, Durham, NC USA}
}

\maketitle

\begin{abstract}
Accurate calibration of control parameters in quantum gates is crucial for high-fidelity operations, yet it represents a significant time and resource challenge, necessitating periods of downtime for quantum computers. Robust Phase Estimation (RPE) \cite{kimmel} has emerged as a practical and effective calibration technique aimed at tackling this challenge. It combines a provably efficient number of control pulses with a classical post-processing algorithm to estimate the phase accumulated by a quantum gate. We introduce Bayesian Robust Phase Estimation (BRPE), an innovative approach that integrates Bayesian parameter estimation into the classical post-processing phase to reduce the sampling overhead. Our numerical analysis shows that BRPE markedly reduces phase estimation errors, requiring approximately $50\%$ fewer samples than standard RPE. Specifically, in an ideal, noise-free setting, it achieves up to a $96\%$ reduction in average absolute estimation error for a fixed sample cost of $88$ shots when compared to RPE. Under a depolarizing noise model, it attains up to a $47\%$ reduction for a fixed cost of $176$ shots. Additionally, we adapt BRPE for Ramsey spectroscopy applications and successfully implement it experimentally in a trapped ion system. 
\end{abstract}


\section{Introduction}
Progress in quantum computing has accelerated in recent years and systems containing tens of qubits now exist in both academia and industry\cite{arute_qs, wang, pino, egan, zhao, ebadi}. Quantum gate fidelities have likewise improved and the field has moved closer towards fault-tolerant quantum computing \cite{preskill}. System calibration has played a pivotal role in the progress up to date. However, current calibration methods continue to be time and resource-intensive processes, requiring up to four hours per day for system sizes of tens of qubits \cite{arute_qs}. This calibration time constitutes downtime during which experiments or applications cannot be run. As quantum computers begin to include hundreds and thousands of qubits the calibration overhead becomes untenable. Novel and time-efficient calibration methods are needed for both single and two-qubit gates.

Robust Phase Estimation (RPE) \cite{kimmel} has established itself as a practical calibration routine for both single-qubit and two-qubit gates. RPE is a non-adaptive parameter estimation protocol which can tolerate state preparation and measurement (SPAM) errors up to a threshold, and exhibits Heisenberg-like scaling with regards to control system pulses. However, this does not always translate to a reduction in overall runtime as the sampling overhead required by the classical post-processing algorithm can become unfavorable \cite{rudinger, meier}.

In this work we propose a novel post-processing algorithm which uses Bayesian parameter estimation instead of the conventional trigonometric algorithm employed in RPE. Through simulation and experiment we show our method, which we call Bayesian Robust Phase Estimation (BRPE), has several advantages over traditional RPE, particularly when the sampling overhead is low and information is limited. In these scenarios, we have found that BRPE not only cuts the sampling overhead by roughly $50\%$ compared to standard RPE methods but also decreases the average absolute error of estimates. Specifically, in an noise-free setting, BRPE reduces estimation error by $96\%$ for a fixed sampling overhead of $88$ shots, and under conditions of depolarizing noise, the reduction is $47\%$ for a fixed cost of $176$ shots. Further, BRPE exhibits improved tolerance to coherent errors during state preparation and measurement operations, which we refer to as coherent SPAM (C-SPAM) errors and measure in radian difference from the desired rotation. In the presence of this type of error, BRPE achieves accurate phase estimates even with C-SPAM errors as high as $0.533$ radians. This is an improvement over traditional RPE, which is limited to C-SPAM errors up to $0.167$ radians. In scenarios where these coherent errors are commutative and directly correlated with the target gate, such as during Rabi frequency calibration, BRPE's algorithm can be slightly modified to extend resilience and BRPE can tolerate C-SPAM errors up to $1.547$ radians. 

This paper is outlined as follows. Section II briefly discusses the two concepts underlying BRPE, namely RPE and Bayesian parameter estimation. Section III provides details on the BRPE calibration and its post-processing algorithm. Section IV includes numerical analysis which explores BRPE and RPE performance using an error-free model, as well as models which include measurement error, depolarizing error and C-SPAM error. In section V BRPE is adapted for Ramsey spectroscopy and is used to experimentally calibrate the carrier frequency for single-qubit gates in a trapped ion quantum computer. Finally, in Section VI future directions are discussed.

\section{Background}

\subsection{Robust Phase Estimation} \label{sec:rpe}
Robust Phase Estimation (RPE) \cite{kimmel} is a protocol which can be used to characterize the rotation angles and axes of a quantum gate. For simplicity we describe the single parameter, single-qubit implementation of RPE. 

RPE defines the target quantum gate as
\begin{align}
    U(\theta) = \cos(\frac{\theta}{2})\sigma_I - i\sin(\frac{\theta}{2})\sigma_X \label{eq:ideal_gate}
\end{align}
where $\sigma_I$ and $\sigma_X$ are Pauli operators, and $\theta$ is the parameter to be estimated. To accurately estimate $\theta$, RPE conducts multiple rounds of experiments, which are indexed by $k=0, 1, \ldots, K$. Each round consists of $M_k$ repetitions of two different experimental sequences. In the first sequence, the qubit is initialized to the $\ket{0}$ state and the unitary, $U(\theta)$, is applied $N(k) = 2^k$ times. The qubit is measured in the $\sigma_Z$ basis which projects it to the $\ket{0}$ or $\ket{1}$ state. A transition probability, $P_{k,a} = a_k/M_k$ is calculated, where $a_k$ is the number of times the qubit was measured in the $\ket{1}$ state when using sequence $1$ in round $k$.  The second sequence initializes the qubit in the $\ket{i}=\frac{\ket{0} + i\ket{1}}{\sqrt{2}}$ Again $U(\theta)$ is applied $N(k)$ times and the qubit is measured in the $\sigma_Z$ basis. A transition probability $P_{k,b} = b_k/M_k$ is calculated, where $b_k$ is the number of times the qubit was measured in the $\ket{1}$ state when using sequence $2$ in round $k$. RPE then uses $P_{k,a}$ and $P_{k,b}$ in a classical post-processing step to generate estimates of $\theta$.

For each round $k$, beginning with $k=0$, the classical algorithm generates a set of possible estimates $\tilde{\Theta}_k$ defined as \cite{russo}
\begin{align}\label{rpe_theta_est}
    \tilde{\Theta}_k \coloneqq \Big \{ &\tilde{\theta}  \;\Big|\; \exists n \in \mathbb{Z} : \nonumber \\ 
    &\tilde{\theta} = \arctan(2P_{k,a}-1, 2P_{k,b} - 1)/N(k) + \frac{2\pi n}{N(k)} \Big \} 
\end{align}
Next, it selects the estimate $\tilde{\theta}_k \in \tilde{\Theta}_k$ which is closest to the estimate found in the previous round, $\tilde{\theta}_{k-1}$. In each round, RPE restricts its estimate to a smaller range such that in the end it finds estimate $\tilde{\theta}_K$ which estimates $\theta$ with error at most $\frac{\pi}{N(K)}$ \cite{kimmel}. As proven in \cite{belliardo}, as long as
\begin{align} \label{rpe_constraint}
    |\tilde{\theta}_k - \theta| \leq \frac{\pi}{3 \cdot 2^{k-1}}
\end{align}
RPE is guaranteed to find an accurate estimate. However, if (\ref{rpe_constraint}) is not satisfied all future rounds of RPE will be incorrect and the calibration routine will fail to find a satisfactory estimate of $\theta$. Recent work has provided consistency tests which can identify if the algorithm has failed at round $k$ \cite{russo}. These consistency tests can be used by experimentalists to identify when RPE results can be trusted. 


Importantly, when gate count (or control system pulses) is the resource of interest, RPE shows Heisenberg scaling up to constant factors \cite{rudinger}. Moreover, RPE is simple to implement as it does not involve entangled states or ancilla qubits, and it is non-adpative, meaning there is no closed-loop, quantum-classical communication or classical optimizer. It has already seen success in experimental trapped ion \cite{rudinger} and superconducting \cite{russo_2} systems. Despite these positive attributes, RPE requires many experimental samples, especially when characterizing noisy systems \cite{meier}.

\subsection{Bayesian Parameter Estimation} \label{subsection_bayes}

By incorporating prior knowledge about a system and repeatedly updating this knowledge, Bayesian parameter estimation has been shown to reduce overall resource costs and calibration run-time for various quantum calibration procedures \cite{bussjaeger, mcmichael, kaubruegger}. Although the implementation details differ depending on the specific calibration, at the heart of Bayesian parameter estimation is Bayes theorem \cite{lee}

\begin{align}
    P(\tilde{\theta} | X) = \frac{P(X | \tilde{\theta})P(\tilde{\theta})}{\int_{\tilde{\theta}} P(X | \tilde{\theta})P(\tilde{\theta})}
\end{align}
where $\tilde{\theta}$ represents the parameters we wish to estimate, $X$ is measured experimental data, $P(X | \tilde{\theta})$ is the likelihood of $X$ given parameters $\tilde{\theta}$, $P(\tilde{\theta})$ is the prior distribution and $P(\tilde{\theta} | X)$ is the posterior distribution. The general procedure for quantum calibration using Bayesian parameter estimation begins with an initial prior distribution , $P_0(\tilde{\theta})$, initial parameter hypothesis, $\tilde{\theta}_0$, and a posterior convergence condition. Typically, $\tilde{\theta}_0$ is determined by sampling $P_0(\tilde{\theta})$. This initial step incorporates prior knowledge about the quantum system and classical control system, and therefore the parameter search begins at an informed starting point. Next, the following iterative process is initiated :

\begin{enumerate}
    \item Run experiments using $\tilde{\theta}_i$ and collect the measured data $X_i$. 
    \item Calculate posterior distribution $P_i(\tilde{\theta} | X_i)$ using Bayes theorem
    \begin{align}
        P_i(\tilde{\theta} | X_i) = \frac{P(X_i | \tilde{\theta})P_i(\tilde{\theta})}{\int P(X_i | \tilde{\theta})P_i(\tilde{\theta})}
    \end{align}
    If $P_i(\tilde{\theta} | X_i)$ satisfies the convergence condition then exit, otherwise continue to Step 3.
    \item Select new experimental parameters $\tilde{\theta}_{i+1}$ using classical optimization techniques. Typically this involves finding an optimal $\tilde{\theta}_{i+1}$ for a given cost function (e.g. cost functions using information entropy, posterior variance, etc.).
    \item Define a new prior distribution equal to the calculated posterior distribution, $P_{i+1}(\tilde{\theta}) = P_i(\tilde{\theta} | X_i)$
    \item Return to Step 1.
\end{enumerate}

Through this process, Bayesian parameter estimation will generate a series of posterior distributions. Ideally, this series converges to a final posterior distribution with an acceptable credible interval thus allowing one to estimate $\tilde{\theta}$ to within some predetermined error.

\section{Bayesian Robust Phase Estimation} \label{sec:brpe}
Although RPE is Heisenberg-limited in control pulses its required sampling overhead significantly increases in the presence of errors \cite{rudinger, russo}. For a single calibration, this overhead may be tolerable and may not substantially affect overall runtime. However, this issue escalates as the number of gates requiring calibration grows, making time increase a limiting factor. Bayesian Robust Phase Estimation (BRPE) addresses this challenge by delivering improved accuracy and precision with fewer samples, effectively enhancing scalability and practicality. 

BRPE follows the same experimental implementation as RPE discussed in section \ref{sec:rpe}, and thus maintains the same Heisenberg-like scaling. It varies in the post-processing of $a_k$ and $b_k$. Rather than calculating transition probabilities $P_{k,a}$ and $P_{k,b}$ it uses each $a_k$ and $b_k$ in Bayesian parameter estimation. 

To begin, BRPE starts with an empirically defined initial prior $P_0(\tilde{\theta})$. For each round $k$, BRPE defines two likelihoods
\begin{align}
    \mathcal{L}_{k,a}(x) &= \cos^2(2^{-k}\tilde{\theta} + (1-x)\frac{\pi}{2}) \label{eq:likelihood1}\\
    \mathcal{L}_{k,b}(x) &= \cos^2(2^{-k}\tilde{\theta} + \frac{\pi}{4} + (1-x)\frac{\pi}{2}) \label{eq:likelihood2}
\end{align}
where $x \in \{0,1\}$. It then generates $4$ intra-round posteriors based on the measured values $a_k$ and $b_k$,

\begin{align}
    P_{k,a,1}(\tilde{\theta}|a_k) &= \frac{P_{k}(\tilde{\theta})\mathcal{L}_{k,a}(1)^{a_k}}{\int_{\tilde{\theta}}P_{k}(\tilde{\theta})\mathcal{L}_{k,a}(1)^{a_k}} \\
    P_{k,a,0}(\tilde{\theta}|a_k) &= \frac{P_{k,a,1}(\tilde{\theta}|a_k)\mathcal{L}_{k,a}(0)^{M_k - a_k}}{\int_{\tilde{\theta}} P_{k,a,1}(\tilde{\theta}|a_k)\mathcal{L}_{k,a}(0)^{M_k - a_k}}\\
    P_{k,b,1}(\tilde{\theta}|b_k) &= \frac{P_{k,a,0}(\tilde{\theta}|a_k)\mathcal{L}_{k,b}(1)^{b_k}}{\int_{\tilde{\theta}}P_{k,a,0}(\tilde{\theta}|a_k)\mathcal{L}_{k,b}(1)^{b_k}} \\
    P_{k,b,0}(\tilde{\theta}|b_k) &= \frac{P_{k,b,1}(\tilde{\theta}|b_k)\mathcal{L}_{k,b}(0)^{M_k - b_k}}{\int_{\tilde{\theta}} P_{k,b,1}(\tilde{\theta}|b_k)\mathcal{L}_{k,b}(0)^{M_k - b_k}}
\end{align}
The last posterior is then used as the prior for the next round, $P_{k+1}(\tilde{\theta}) = P_{k,b,0}(\tilde{\theta}|b_k)$. This process is repeated for all $k$ rounds until a final posterior $P_{n}(\tilde{\theta}|a_0, b_0, \ldots, a_{n-1}, b_{n-1})$ is generated.

After processing all $k$ rounds, BRPE uses a maximum \textit{a posteriori} probability (MAP) estimate on the final posterior distribution to find $\tilde{\theta}$. 

Like RPE, BRPE's estimation accuracy degrades as errors increase, albeit more slowly than RPE for most types of errors (see Section \ref{sec:simulations}). Accuracy degradation typically stems from a multi-modal posterior distribution in which the difference between several local maxima is too small to definitely distinguish an accurate $\tilde{\theta}$. If the total variance of the local maxima is less than the desired precision in the estimate then the $\tilde{\theta}$ corresponding to any of these local maxima will suffice. However, if the variance is greater it becomes more difficult to determine an accurate $\tilde{\theta}$, with the difficulty being proportional to the variance. To account for these scenarios, the last step of BRPE is the calculation of a heuristic confidence score 
\begin{align}\label{confidence_score}
    C(D, \sigma, \sigma_{max}) &= \frac{1}{1 + \exp(
        -5(f(D, \sigma, \sigma_{max})-\frac{3}{4})} \\
    f(D, \sigma, \sigma_{max}) &= \frac{3D}{2} + \frac{1}{4}(\frac{1}{1+\frac{1}{10}\exp(\frac{1}{4}(\sigma - \sigma_{max}))})
\end{align}
where $D = \max\limits_{\tilde{\theta}} P_{n}(\tilde{\theta}|a_0, b_0, \ldots, a_{n-1}, b_{n-1})$ is the MAP estimate on the final posterior distribution, $\sigma$ is the standard deviation of the local maxima in the final posterior distribution, and $\sigma_{max}$ is an empirically selected threshold. This score encodes the conditions described above and quantifies the confidence in the final estimate. In section \ref{sec:experimental_results} we demonstrate the utility of the confidence score and examine how it can assist in the implementation of BRPE. We note that this score can be further modified to incorporate other conditions and constraints. 

\begin{figure*}[hbt!]
     \centering
     \begin{subfigure}[b]{\textwidth}
         \centering
         \includegraphics[width=0.7\textwidth]{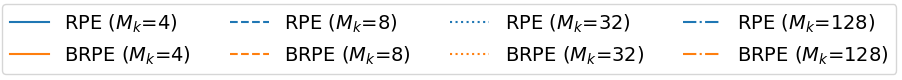}
     \end{subfigure}
     \begin{subfigure}[b]{0.465\textwidth}
         \centering
         \includegraphics[width=\textwidth]{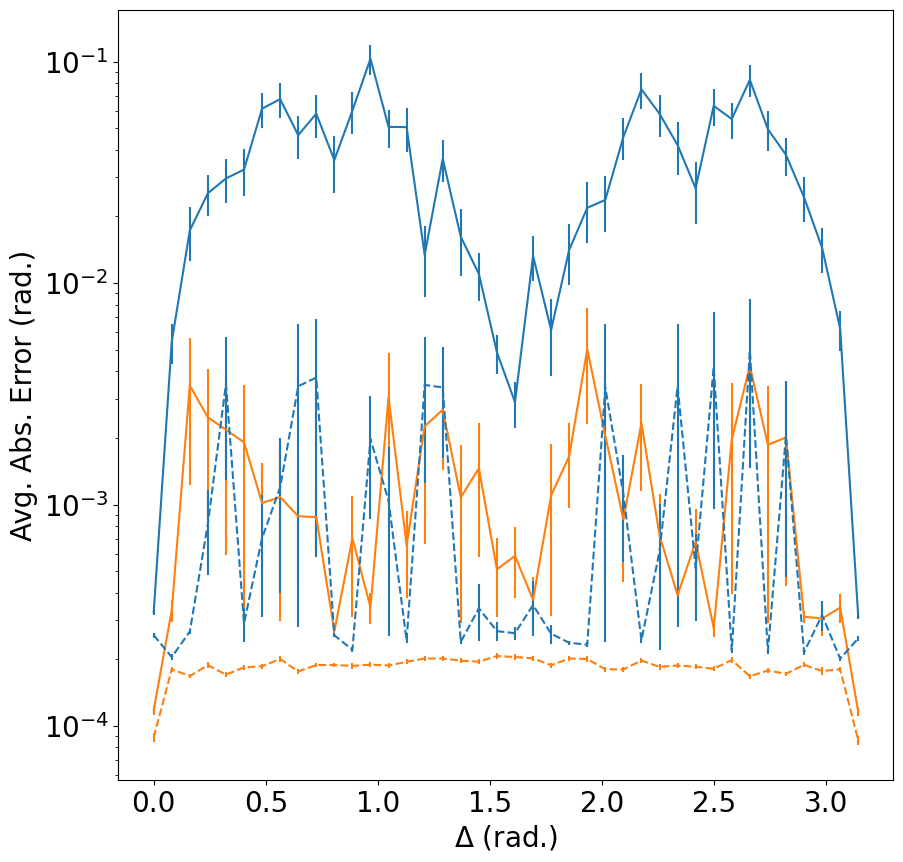}
         \caption{}
         \label{fig:error_free_low_mk}
     \end{subfigure}
     \centering
     \begin{subfigure}[b]{0.5\textwidth}
         \centering
         \includegraphics[width=\textwidth]{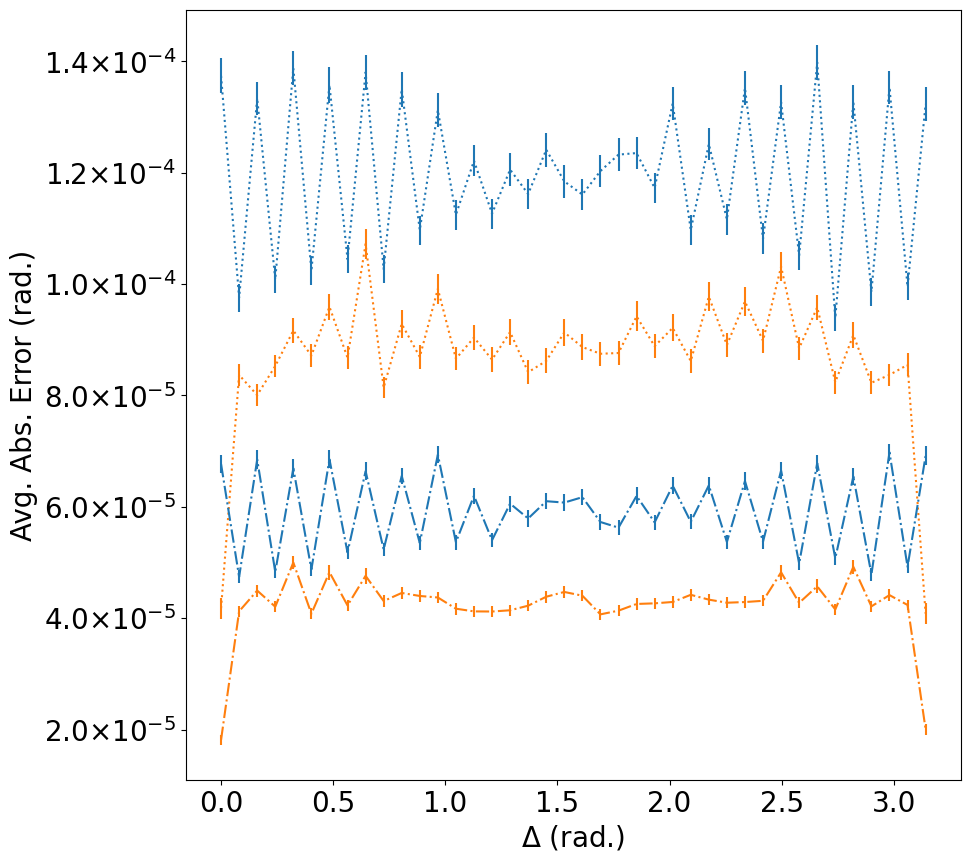}
         \caption{}
         \label{fig:error_free_high_mk}
     \end{subfigure}
        \caption {Plots of $\bar{\varepsilon}_{\Delta_j}$ for RPE and BRPE in an error free environment using different shot counts. Averages are calculated using $1000$ independent trials. Error bars represent standard error. a) When $M_k =4$ and $M_k=8$ BRPE results in a $96.08\% \pm 4.34\%$ and $85.08\% \pm 17.5\%$ decrease in average absolute error, respectively, when compared to RPE. b) BRPE outperforms RPE for all values of $M_k$. RPE requires approximately twice as many shots to match BRPE performance. As shot count increases we see convergence of BRPE and RPE performance. }
        \label{fig:error_free}
\end{figure*}

\section{Numerical Analysis} \label{sec:simulations}

In this section, we assess the accuracy and precision of both BRPE and RPE in estimating phases under various conditions through classical simulations. Additionally, we explore how the estimation errors scale with the number of shots for each method. We have made further statistical analysis beyond the results reported here available publicly \cite{extradata}.

\subsection{Performance Analysis Across Different Error Conditions } \label{sec:performance_analysis}
The accuracy and precision of BRPE and RPE are affected by the operational enviornment. Here, we investigate their performance in an error-free environment as well as error-prone environments characterized by measurement errors, depolarizing errors, and C-SPAM errors. 

Each simulation assumes the ideal gate of $U(\theta)$ (\ref{eq:ideal_gate}) is instead implemented as $U(\Theta)$ where $\Theta = \theta + \Delta$, and $\Delta$ represents an unintentional deviation in phase. BRPE and RPE are then used to find an estimate $\tilde{\Delta}$ of the deviation. Simulations are performed across a range of manufactured offsets $\Delta_j \in \{\frac{\pi \cdot j}{d} \;|\; j=0,\ldots,d \; ; \;d \in \mathcal{Z}^+ \}$ to asses protocol performance across a broad spectrum of phases.

Further, to quantify performance of each protocol we simulate $N$ independent trials of BRPE and RPE for each offset, $\Delta_j$, and calculate the asbolute error in the estimation of each trial $\varepsilon_{i,\Delta_j} = |\tilde{\Delta}_i - (\Theta - \theta)|$, $i=0,\ldots,N-1$. We then calculate the average absolute error, $\bar{\varepsilon}_{\Delta_j}$ at each offset. 
\begin{align}
    \bar{\varepsilon}_{\Delta_j} &= \frac{\sum_{i=0}^{N-1} \varepsilon_{i, \Delta_j}}{N} 
\end{align}

To compare protocols we calculate the average absolute error over all $\Delta_j$
\begin{align}
    \bar{\varepsilon} = \frac{\sum_{i=0}^{d} \bar{\varepsilon}_{\Delta_i}}{d+1}
\end{align}
and use this metric to compare BRPE and RPE.  

Finally, we define protocol failure to occur when average error is above a threshold, $\bar{\varepsilon}_{\Delta_j} > \bar{\varepsilon}_{th}$. To determine this threshold we find the maximum error in estimation which results in a gate fidelity of $\mathcal{F} \geq 0.9995$. 

Throughout this section we use $\theta = \frac{\pi}{2}$, $N=1000$, $d = 40$, $\bar{\varepsilon}_{th} = 0.02$ and we fix the number of rounds to $K=11$ for both protocols.

\subsubsection{Error Free} \label{sec:error_free}

This section evaluates the performance of BRPE and RPE in an error-free environment, particularly focusing on the relationship between their performance and the shot count, denoted as $M_k$. It is observed that BRPE generally surpasses RPE in terms of performance when both methods are applied with an equivalent number of shots, as depicted in Figures \ref{fig:error_free_low_mk} and \ref{fig:error_free_high_mk}. A notable disparity is evident at lower shot counts. Specifically, for $M_k=4$, RPE's average error ($\bar{\varepsilon} = 3.5 \times 10^{-2} \pm 2.5 \times 10^{-2}$) surpasses the threshold $\bar{\varepsilon}_{th} = 0.02$, indicating its likely failure in accurately calibrating the target gate to the desired specification. In contrast, BRPE maintains an average error below this threshold at the same shot count, with $\bar{\varepsilon} = 1.4 \times 10^{-3} \pm 1.2 \times 10^{-3}$. Although this performance gap diminishes with increasing shot counts, RPE generally requires about twice as many shots as BRPE to achieve comparable results. 

\begin{figure*}[hbt!]
     \centering
     \begin{subfigure}[b]{0.7\textwidth}
         \centering
         \includegraphics[width=\textwidth]{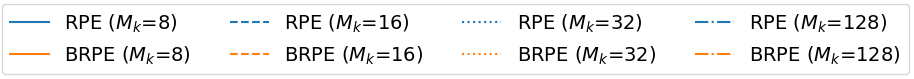}
     \end{subfigure}
     \begin{subfigure}[b]{0.3\textwidth}
         \centering
         \includegraphics[width=\textwidth]{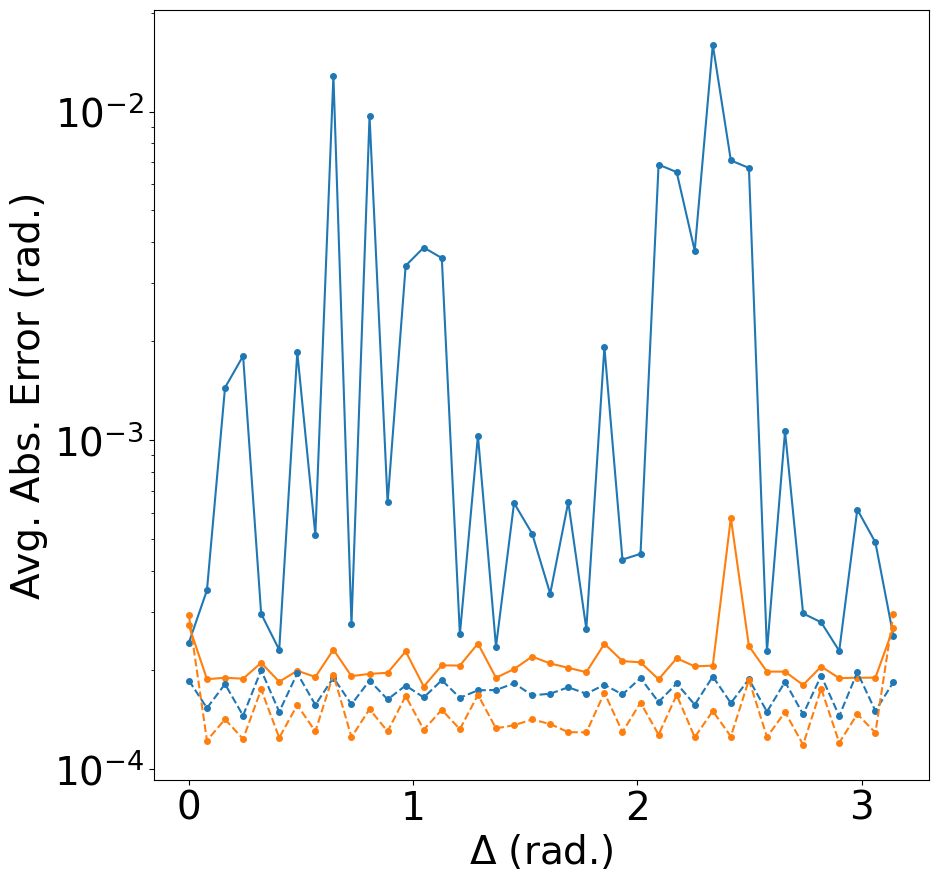}
         \caption{}
         \label{fig:read_error_p1_low_count}
     \end{subfigure}
     \centering
     \begin{subfigure}[b]{0.33\textwidth}
         \centering
         \includegraphics[width=\textwidth]{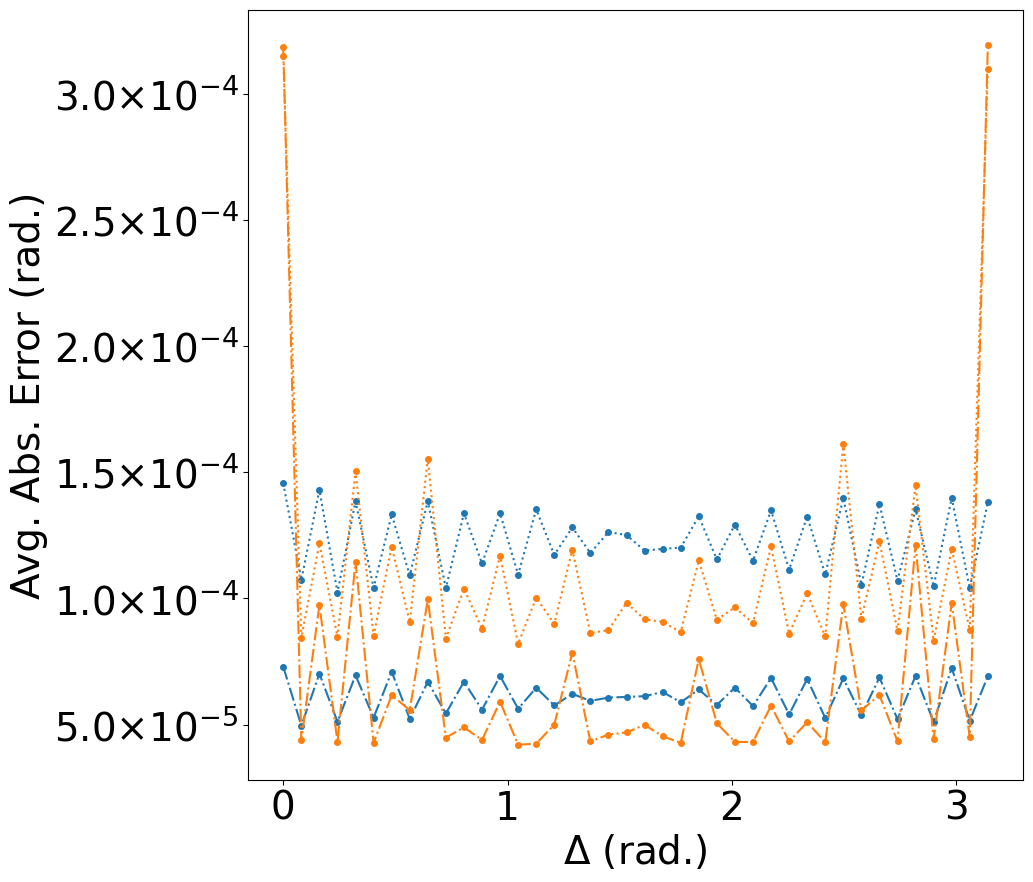}
         \caption{}
         \label{fig:read_error_p1_high_count}
     \end{subfigure}
     \centering
     \begin{subfigure}[b]{0.33\textwidth}
         \centering
         \includegraphics[width=\textwidth]{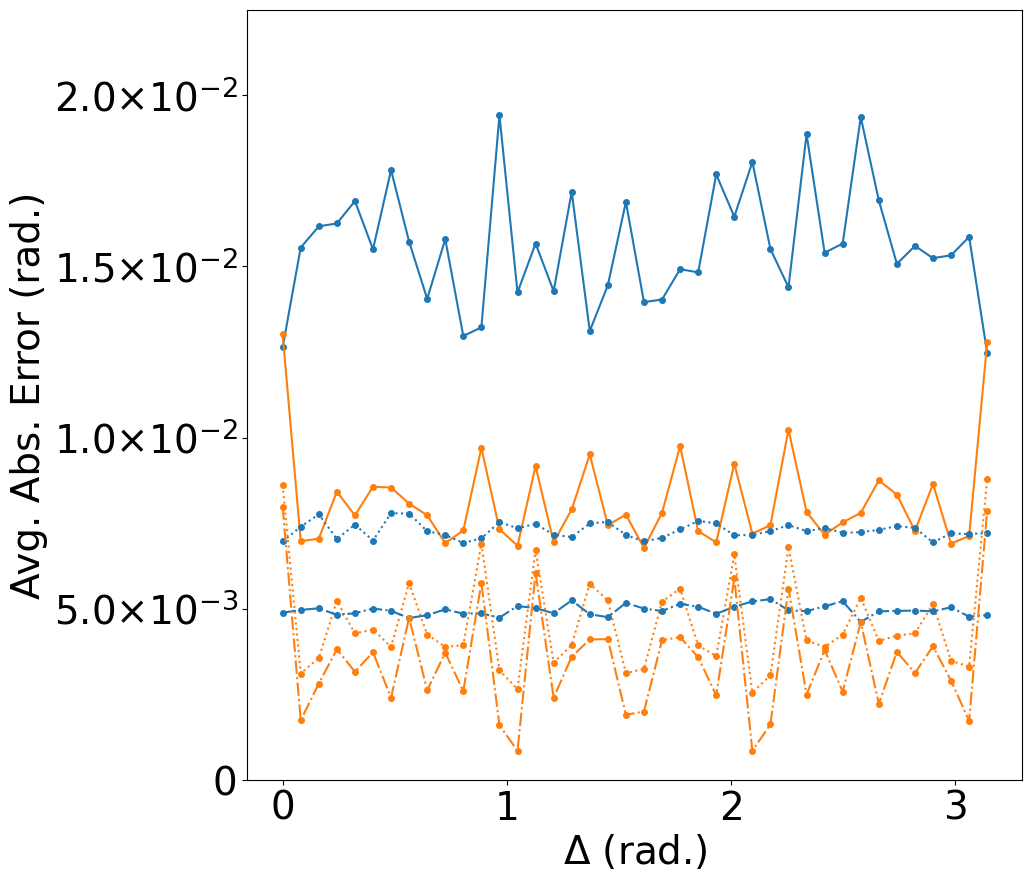}
         \caption{}
         \label{fig:dep_errors}
     \end{subfigure}
        \caption{Performance of RPE and BRPE for $0 \leq \Delta_j < \pi$ in the presence of measurement error and depolarizing error with varying shot count. Averages are calculated using $1000$ independent trials. a) When measurement error is set to $p=0.01$ and shot count $M_k \leq 16$ BRPE yields estimates with average absolute error substantially below that of RPE. b) Increasing shot count while maintaining a measurement error of $p=0.01$ provides a greater performance boost for RPE than BRPE. BRPE continues to yield the overall minimum lowest average absolute error across all values of $\Delta_j$, but inconsistently outperforms RPE. BRPE outliers at boundaries can be mitigated by shifting search space over $\Delta_j$ (See Section \ref{sec:read_error}). c) In the presence of a $1\%$ depolarizing error BRPE results in lower average absolute error than RPE. As shot count increases RPE performance begins to the converge to that of BRPE.}
        \label{fig:read_errors}
\end{figure*}

\subsubsection{Measurement Error} \label{sec:read_error}

We next investigate the performance of BRPE and RPE under the influence of measurement error. This error is modeled using a post-processing step in which we pass the measured, ideal expectation value through a binary symmetric channel. Designating $p$ as the crossover probability we evaluate performance using $p \in \{0.005, 0.01, 0.015, 0.02, 0.025, 0.03\}$ and shot count $M_k \in \{8, 16, 32, 64, 128\}$. In this section we use $M_k=8$ as the lower limit of our shot count. This choice is informed by our findings in Section \ref{sec:error_free}, where we determined that a minimum of $8$ shots is required to ensure the successful operation of both protocols.

In scenarios with lower shot counts ($M_k \leq 16$), BRPE demonstrates superior performance over RPE across all measurement error probabilities. This is clearly illustrated in Fig. \ref{fig:read_error_p1_low_count} where BRPE's advantage is well pronounced at $p=0.01$. The same advantage is observed at $M_k=8$ and $p=0.03$, where RPE achieves an average absolute error ($\bar{\varepsilon}$) of less than $3.7\times 10^{-3} \pm 3.3\times 10^{-3}$, corresponding to a gate fidelity error ($1-\mathcal{F}$) of approximately $1.74\times 10^{-6}$. In contrast, BRPE achieves a significantly lower $\bar{\varepsilon}$ of less than $2.51\times 10^{-4} \pm 8.2\times 10^{-5}$, with a corresponding fidelity error of about $8.33\times 10^{-9}$, underscoring BRPE's enhanced accuracy in low-shot scenarios, even under high measurement errors.

For applications requiring higher gate fidelity or reduced average absolute error, an increase in shot count becomes necessary. Fig. \ref{fig:read_error_p1_high_count} presents the performance of both protocols at higher shot counts for $p=0.01$. As more shots are utilized to minimize the average absolute error, the performances of BRPE and RPE start to converge. Notably, in high-error-rate, high-shot-count scenarios, RPE begins to outperform BRPE. For example, at $M_k=64$ and $p=0.03$, RPE yields an average absolute error of $9.08 \times 10^{-5} \pm 8.9 \times 10^{-6}$ whereas BRPE produces estimates with average absolute error of $1.19 \times 10^{-4} \pm 8.6 \times 10^{-5}$.  However, it is important to note that this shift in performance advantage occurs when the target gate fidelity error is on the order of $10^{-9}$.

In Fig. \ref{fig:read_error_p1_high_count}, it is noteworthy that we observe significant values for $\bar{\varepsilon}_{\Delta_j}$ at both $\Delta_j = 0$ and $\Delta_j=\pi$. In the presence of measurement errors, the default implementation of BRPE performs less effectively than RPE when $\Delta_j \lessapprox 1 \times 10^{-3}$ and $M_k > 8$.

This performance discrepancy arises because when $\Delta_j$ is much smaller than $1 \times 10^{-3}$, increasing the value of $k$ does not provide substantial new information. As a result, the Bayesian process lacks sufficient additional data to improve the posterior distribution effectively. Due to the periodicity of our likelihoods the same discrepancy arises when $\Delta_j \approx \pi$. 

However, a solution can be found by introducing an artificial offset, denoted as $U(\theta + \delta)$, where $\delta > 1 \times 10^{-3}$. By accounting for this offset when calculating $\tilde{\theta}$, we can effectively restore BRPE's performance, even in scenarios where $\Delta_j$ is significantly less than $1 \times 10^{-3}$.

\begin{figure*}[hbt!]
     \centering
     \begin{subfigure}[b]{\textwidth}
         \centering
         \includegraphics[width=0.7\textwidth]{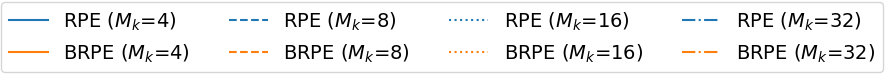}
     \end{subfigure}
     \begin{subfigure}[b]{0.33\textwidth}
         \centering
         \includegraphics[width=\textwidth]{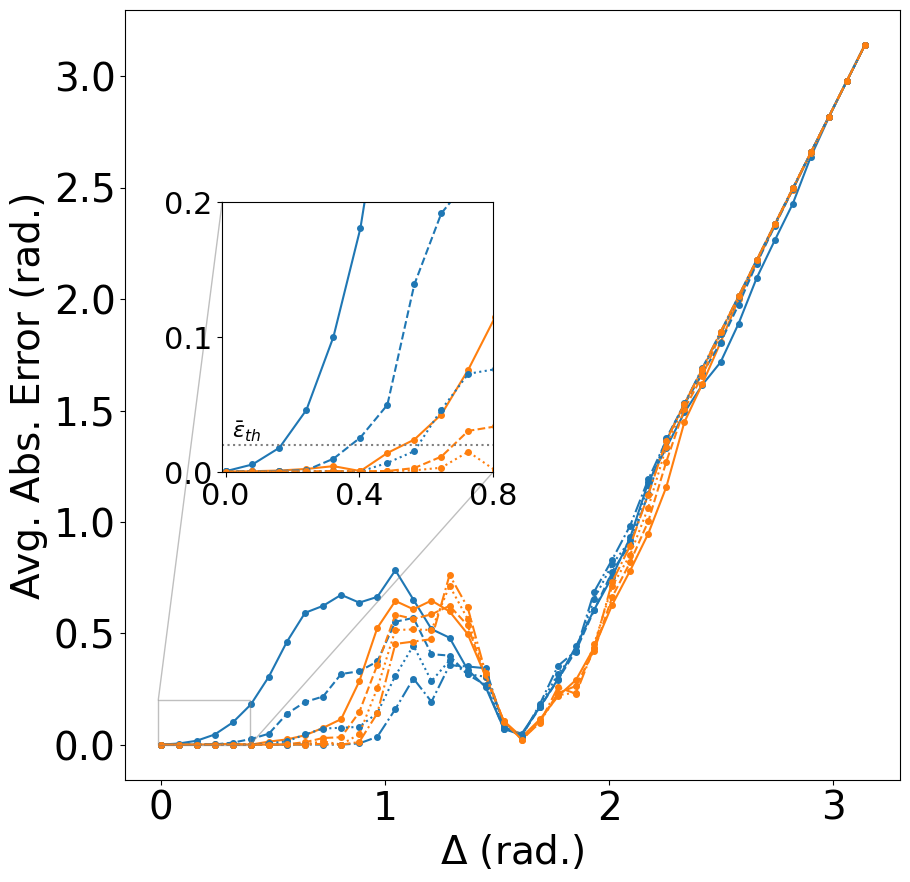}
         \caption{}
         \label{fig:corr_coh}
     \end{subfigure}
     \centering
     \begin{subfigure}[b]{0.33\textwidth}
         \centering
         \includegraphics[width=\textwidth]{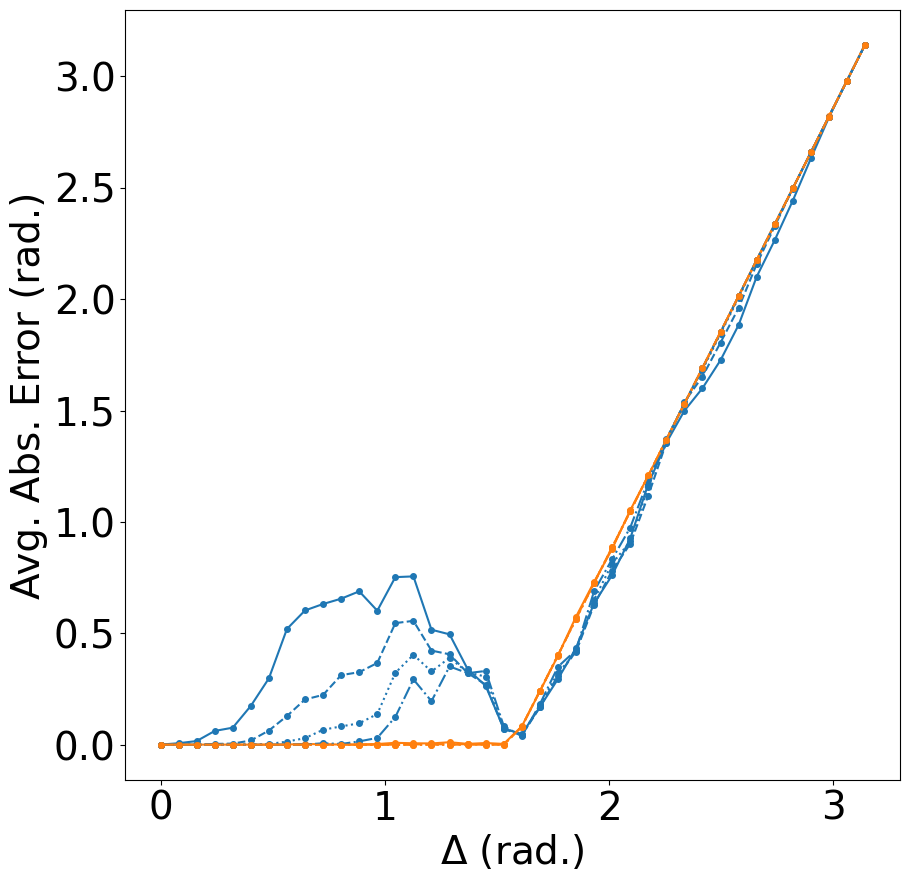}
         \caption{}
         \label{fig:comm_propto}
     \end{subfigure}
        \caption{Performance of RPE and BRPE for $0 \leq \Delta_j < \pi$ in the presence of C-SPAM errors. C-SPAM errors are modeled as coherent overrotations along the same axis as the target gate. Averages are calculated using $1000$ independent trials. Notably, the magnitude of these errors is correlated with the error present in the target gate itself. a) BRPE is implemented using likelihoods defined in eqs. \ref{eq:likelihood1} and \ref{eq:likelihood2}. When evaluated across the entire range of $\Delta$'s, BRPE and RPE performance varies. Notably, both protocols fail to find reasonable estimates beyond $\Delta > 0.925$. However, with C-SPAM errors $\leq 0.839$ radians — a significant error margin — BRPE outperforms RPE in terms of lower average absolute error when using $M_k \leq 16$ shots. It also shows greater C-SPAM error tolerance, evident from the intersection points with the $\bar{\varepsilon}_{th}$ threshold (see inset).  b) When BRPE likelihoods are modified to eqs. \ref{eq:likelihood3} and \ref{eq:likelihood4} performance gains are substantially greater. Average absolute error is further reduced by using the tailored likelihoods and BRPE achieves tolerance to C-SPAM errors up to $\pi/2$
        . }
        \label{fig:spam_errors}
\end{figure*}

\subsubsection{Depolarizing Error}\label{sec:depolarizing_error}

We next look at the performance of RPE and BRPE when our target gate is subject to a depolarizing error channel. Specifically, we use the error channel 
\begin{align}
    \mathcal{E}(\rho) = (1-p)\rho + \frac{p}{3}(\sigma_X\rho \sigma_X +\sigma_Y\rho \sigma_Y + \sigma_Z\rho \sigma_Z)
\end{align}
where $\rho = \ket{\psi}\bra{\psi}$ is the density matrix representation of a quantum state and $p$ is the probability of error.

Fig. \ref{fig:dep_errors} presents a simulation of both BRPE and RPE in an environment with a $1\%$ depolarizing error. In this scenario, both protocols effectively produce estimates with errors below the threshold $\bar{\varepsilon}_{th}$ for shot counts $M_k \geq 8$. Notably, BRPE consistently outperforms RPE by achieving a lower average absolute error across various $M_k$ values. The reduction in average error achieved by BRPE compared to RPE varies significantly, ranging from approximately $29.93\% \pm 33.21\%$ at $M_k=128$ to a more pronounced $47.78\% \pm 10.73\%$ at $M_k=8$.

The analysis is further extended to a scenario with a higher, $3\%$ depolarizing error. In this case, BRPE continues to provide more precise estimates than RPE. The improvement in average absolute error with BRPE over RPE spans from $30.78\% \pm 30.93\%$ at $M_k=128$ to $46.85\% \pm 10.15\%$ at $M_k=8$. However, a notable distinction arises at this higher error rate: both BRPE and RPE require an increased number of shots to maintain performance below the failure threshold. Specifically, while BRPE maintains effectiveness with a minimum of $M_k \geq 8$ shots, RPE requires a significantly higher shot count of $M_k \geq 64$ to stay under the threshold, highlighting BRPE's superior robustness in higher-error conditions.

We again see BRPE demonstrates heightened sensitivity to depolarizing errors when $\Delta_j$ is on the order of $1 \times 10^{-3}$, for analogous reasons to its sensitivity in the presence of measurement errors. A similar mitigation technique as discussed in section \ref{sec:read_error} can be used.


\subsubsection{C-SPAM Error}

Finally we explore BRPE and RPE performance in the presence of SPAM errors. Given the target gate described in (\ref{eq:ideal_gate}) and the measurement bases described in Section \ref{sec:rpe}, modeling SPAM as a depolarizing channel on the state preparation and measurement gates will yield results similar to Section \ref{sec:depolarizing_error}. Instead we explore the impact on performance when SPAM errors are modeled as coherent overrotations in the state preparation and measurement gates with the amount of overrotation directly correlating with the error in the target gate, which we refer to as C-SPAM errors. Further we note preparation of the $\ket{i}$ state is implemented using an x-axis rotation and thus utilizes the same axis as the target gate defined in (\ref{eq:ideal_gate}). This model aligns with real-world scenarios where state preparation gates, measurement gates, or both, utilize rotations along the same axis as the target gate. One common example of this is Rabi frequency calibrations.

In Fig. \ref{fig:corr_coh} we see both BRPE and RPE yield accurate estimations up to an overrotation threshold, $\Delta_{th}$,
\begin{align}
    \Delta_{th} = \max \{ \Delta_j | \bar{\varepsilon}_{\Delta_j} < \bar{\varepsilon}_{th} \}
\end{align}
after which each protocol fails to provide an estimate with error below our error threshold, $\bar{\varepsilon}_{th}$. 

At lower shots counts, BRPE has both lower average error (calculated up to $\Delta_{th}$) and higher $\Delta_{th}$. For example, at $M_k=16$ BRPE yields $\bar{\varepsilon} = 1.9 \times 10^{-3} \pm 4.2\times 10^{-3}$ up to $\Delta_{th} = 0.839$ whereas RPE results in $\bar{\varepsilon} = 2.9 \times 10^{-3} \pm 5.1\times 10^{-3}$ up to $\Delta_{th} = 0.576$. This changes when $M_k \geq 32$ and RPE begins to provide better estimates at higher overrotation when compared BRPE. However, it is important to note that $\Delta \approx 0.839$ represents a sizeable overrotation error and we therefore expect BRPE to be the more efficient protocol for the majority of calibration scenarios that align with our model.

\begin{figure*}[hbt!]
    \centering
    \includegraphics[width=0.4\textwidth]{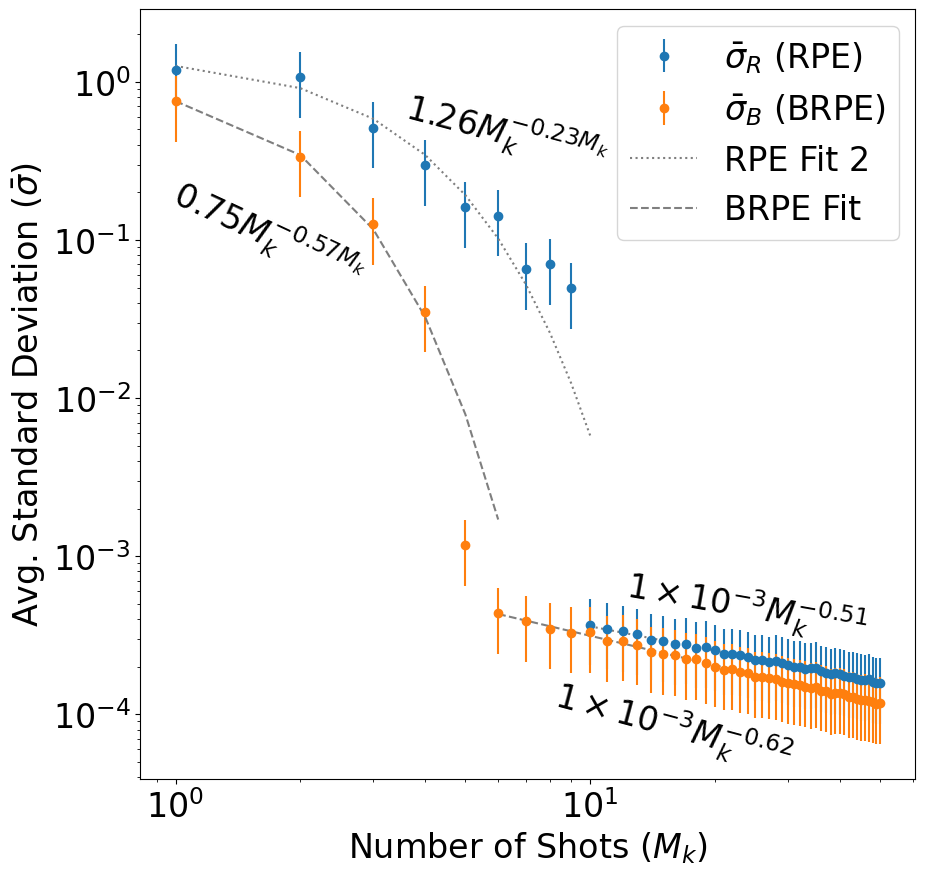}
    \caption{The relationship between shot count, $M_k$, and estimate error, quantified by standard deviation, was evaluated for both BRPE and RPE. Here, the phase offset was fixed at $\Delta=0.201$ radians and both protocols were used to find estimates of $\Delta$. We computed the standard deviation across $1000$ independent simulations for each specific shot count. This process was replicated $5$ times, and each point in the figure represents the average standard deviation calculated as $\bar{\sigma} = \sqrt{\sigma_1^2 + \ldots + \sigma_5^2}$. We then heuristically determined fits by identifying functions that most accurately represented the observed data. The error scaling behavior of BRPE and RPE can be separated into two distinct phases. In the initial phase, characterized by low shot counts, both protocols exhibit a rapid reduction in error due to the fast rate of information gain as shot count increases. Specifically, BRPE scaling can be approximated as $0.76M_k^{-0.57M_k}$, while RPE follows a scaling approximated by $1.26M_k^{-0.25M_k}$. At higher shot counts, specifically $M_k=6$ for BRPE and $M_k=9$ for RPE, the error scaling transitions to exponential. In the shot count range from $M_k=6$ to $M_k=9$, BRPE demonstrates significant performance improvements over RPE. While BRPE shows substantial performance improvements in the low shot count regime, its advantages become less pronounced at higher shot counts. Furthermore, the error bars representing standard error indicate that these improvements are not statistically significant in this range. }
    \label{fig:comparing_std}
\end{figure*}

Furthermore, if calibration is needed when $\Delta > 0.839$ BRPE can be minimally adapted to substantially outperform RPE. Specifically, we can modify our likelihoods such that eqs. \ref{eq:likelihood1} and \ref{eq:likelihood2} become
\begin{align}
    \mathcal{L}_{k,a}(x) &= \cos^2(2^{-k}\tilde{\theta} + (1-x)\frac{\pi}{2}) \label{eq:likelihood3}\\
    \mathcal{L}_{k,b}(x) &= \cos^2(2^{-(k+1)}\tilde{\theta} + (1-x)\frac{\pi}{2}) \label{eq:likelihood4}
\end{align}
These updated likelihoods more accurately capture our model and allow BRPE to not only provide estimates with lower average errors but also extend $\Delta_{th}$ as seen in Fig. \ref{fig:comm_propto}. When using eqs. \ref{eq:likelihood3} and \ref{eq:likelihood4} as likelihoods, BRPE generates estimates with $\bar{\varepsilon}=1.10 \times 10^{-4} \pm 5.9\times 10^{-5}$ and $\Delta_{th}=1.55$ at $M_k=32$. Noteworthy, this adapted version of BRPE achieves $\Delta_{th} \approx \pi/2 $ for all values of $M_k$. 

While it is possible to adapt RPE for this specific model, the flexibility and adaptability of BRPE stand out as notable advantages. BRPE facilitates the seamless incorporation of various likelihoods by abstracting and encapsulating them within the code, effectively decoupling them from the overarching algorithmic control flow. This abstraction feature provides a modular, plug-and-play component within the BRPE algorithm, enabling practitioners to effortlessly substitute different likelihoods to accommodate adjustments to the system model. In contrast, modifying RPE necessitates alterations throughout the algorithm since the model lacks an isolated representation in RPE.

\subsection{Error Scaling}
As discussed in Section \ref{sec:performance_analysis}, BRPE and RPE display notable performance differences in scenarios with low shot counts. This pattern persists both in error-free environments and in the presence of errors. To further investigate the enhanced performance of BRPE in the low shot count domain, we analyzed how standard deviations for both methods evolve with increasing shot counts. We examined 40 different phase offsets ranging from $0$ to $\pi/4$. For each offset, we simulated both BRPE and RPE across varying shot counts $1 \leq M_k \leq 50$. Each offset-shot count combination underwent $1000$ independent simulations, referred to as a trial. We conducted $5$ such trials per combination and calculated the aggregated standard deviation and standard error for each. 

Figure \ref{fig:comparing_std} illustrates trends consistent across all 40 offsets, with the exception of $\Delta=0$, where both methods demonstrated comparable performance. We numerically fit functions to two distinct regimes, allowing us to estimate the scaling parameters for each protocol. In the initial regime, characterized by very low shot counts, both protocols display rapid error reduction. This marked decrease in error is attributed to the minimal information initially available and the substantial information gains as shot counts increase. In this regime, BRPE surpasses RPE due to its improved scaling, which is approximated by the function $0.76M_k^{-0.57M_k}$. In contrast, RPE exhibits scaling approximated by $1.26M_K^{-0.25M_k}$. Both of these fits were determined through heuristic methods, where we identified functions that most accurately represented the observed data. 

Transitioning to higher shot counts, beyond $M_k=6$ shots for BRPE and $M_k=9$ shots for RPE, the scaling of each protocol shifts to exponential. Specifically, BRPE scales approximately as $1\times10^{-3}M_k^{-0.62}$, and RPE as $1\times 10^{-3}M_k^{-0.51}$. In this regime, BRPE demonstrates marginally better error reduction than RPE. However, the standard errors represented by the error bars suggest that these improvements are not statistically significant. 

Notably, for shot counts between $6$ and $10$, the standard deviation of BRPE remains around $10^{-3}$, significantly lower than RPE's which is closer to $10^{-1}$. This improvement aligns with the performance enhancements observed for $M_k=8$, as detailed in section \ref{sec:performance_analysis}.

\section{Example Implementation of BRPE on a Trapped Ion Quantum Computer}
To validate the functionality of BRPE, we conducted experiments on a trapped ion quantum computer. Furthermore, to illustrate the versatility of RPE and, by extension, BRPE, we employed BRPE to implement an enhanced variant of Ramsey spectroscopy \cite{ramsey}. This application is instrumental in calibrating the carrier frequency of the control laser used to implement a single-qubit gate on a trapped ion quantum computer. 

Before presenting the results of our experiments we briefly discuss how BRPE can be adapted to implement Ramsey spectroscopy.

\subsection{Adapting BRPE for Ramsey Spectroscopy}

Ramsey spectroscopy is an often used technique to calibrate control laser frequency, $\omega_L$, in trapped ion systems when the laser frequency is offset from the transition frequency, $\omega_0$, of a target ion. The interaction Hamiltonian, $H_I$, for this laser-ion system is
\begin{align}
    H_I &= \frac{\Omega}{2}(\cos(\varphi)\sigma_X + \sin(\varphi)\sigma_Y) + \frac{\delta}{2}\sigma_Z
\end{align}
where $\Omega$ is the Rabi frequency, $\varphi$ is the laser phase and $\delta = \omega_L - \omega_0$ is the detuning from the resonance frequency. From this Hamiltonian we can see  the laser offset results in an unwanted $\sigma_Z$ rotation. Ramsey spectroscopy is routinely used to detect and correct for the detuning parameter, $\delta$. Similarly, BRPE can also be used effectively to determine $\delta$.

To accomplish this, we use target gate $U=\sigma_I$ and treat the $\sigma_Z$ rotation which comes from the laser frequency offset as the coherent overrotation of interest as discussed in section \ref{sec:brpe} and simulated in section \ref{sec:simulations}. Next we define a search space $[-\delta_{max}, \delta_{max}]$ where $\delta_{max}$ is an empirically selected absolute maximum expected laser frequency offset. Likelihoods are modified to be
\begin{align}
    \mathcal{L}_{k,a}(x) &= \cos^2(\tilde{\delta}\tau_k + (1-x)\frac{\pi}{2}) \\
    \mathcal{L}_{k,b}(x) &= \cos^2(\tilde{\delta}\tau_k + \frac{\pi}{4} + (1-x)\frac{\pi}{2})
\end{align}
with
\begin{align}
    \tau_k &= \frac{2^{k-1}}{\delta_{max}} \; , \; k = 0, \ldots, K \\
    K &= \lfloor \log (2\delta_{max}\beta) \rfloor
\end{align}
where $\tau_k$ is a Ramsey spectroscopy wait time and $\beta < T_2^*$ a heuristically selected parameter. This selection of $\tau_k$ ensures the period of the initial likelihoods $\mathcal{L}_{0, a}(x)$ is $T_0 = \pi$ and the final wait time, $\tau_K \leq \beta$. 

\begin{figure*}[hbt!]
     \centering
     \begin{subfigure}[b]{0.45\textwidth}
         \centering
         \includegraphics[width=\textwidth]{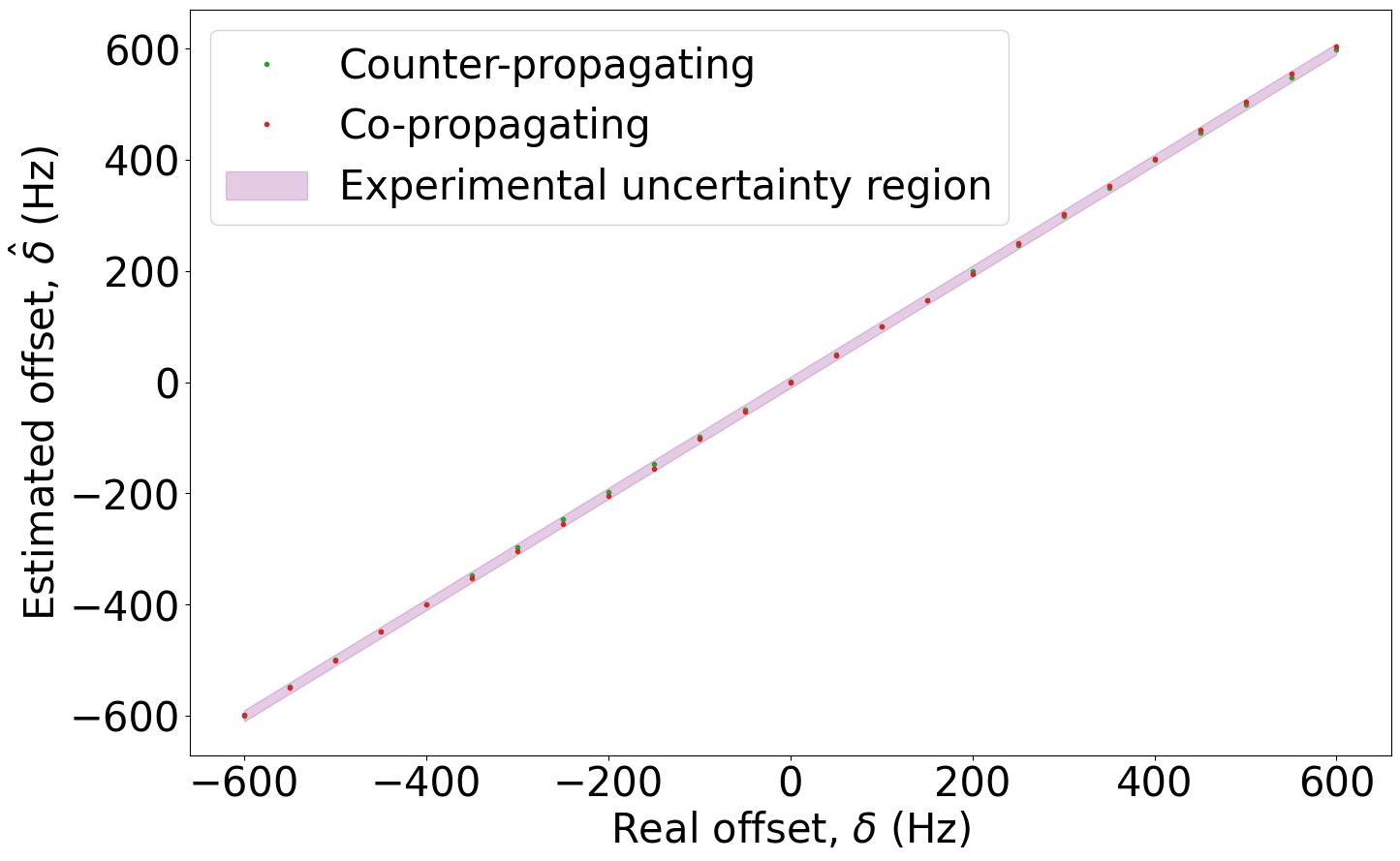}
         \caption{}
         \label{fig:brpe_experiment}
     \end{subfigure}
     \centering
     \begin{subfigure}[b]{0.48\textwidth}
         \centering
         \includegraphics[width=\textwidth]{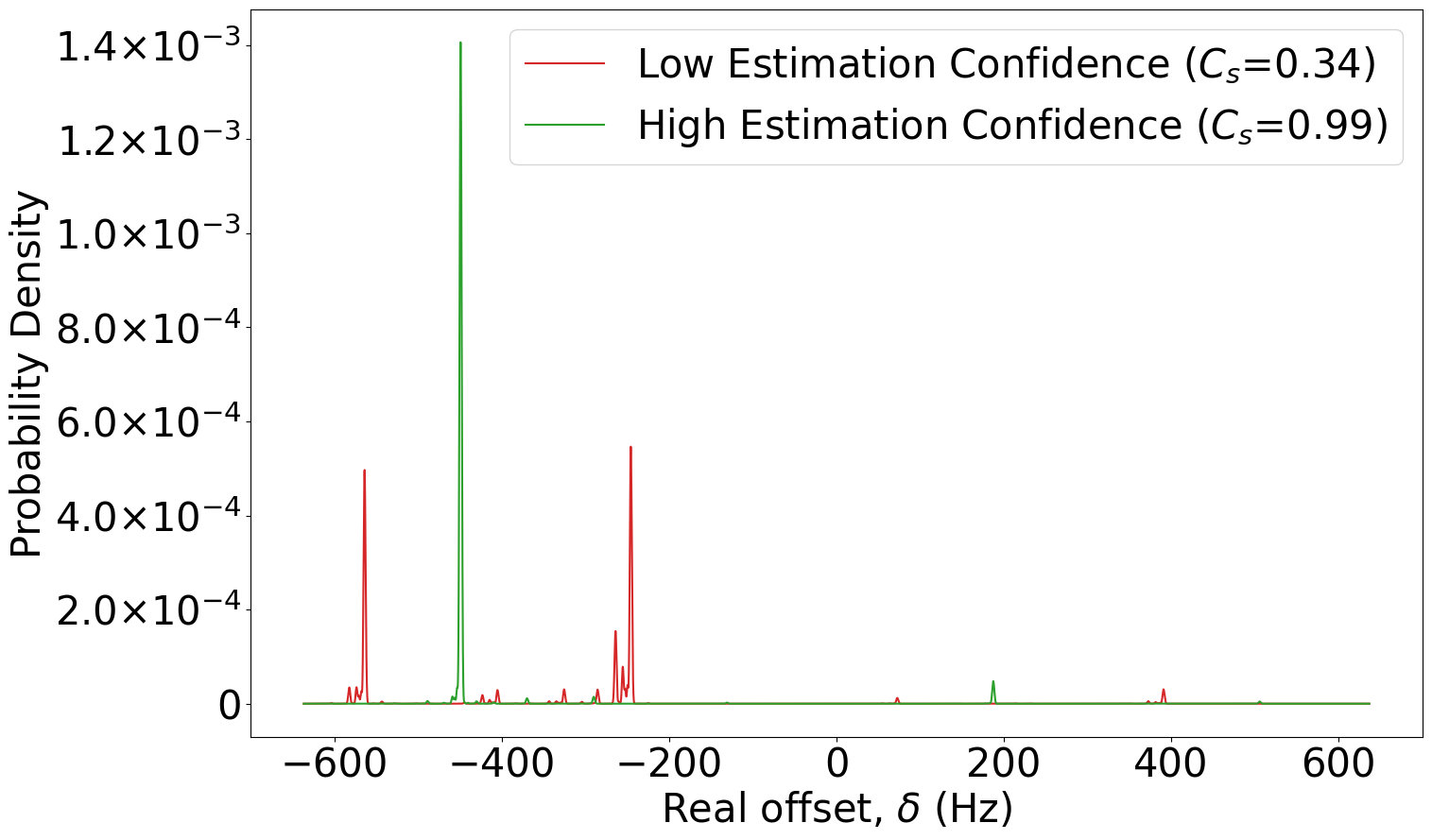}
         \caption{}
         \label{fig:brpe_confidence}
     \end{subfigure}
        \caption{a) The accuracy and precision of BRPE was evaluated in an experimental trapped ion system. For both counter-propagating and co-propagating Raman transitions, the laser frequencies, $\omega_L$, were artificially offset from the transition frequency of the qubit, $\omega_0$, by an amount $\delta$ such that $\omega_L = \omega_0 + \delta$. BRPE was then used to find an estimate, $\tilde{\delta}$, of the offset. When applied to co-propagating Raman transitions, BRPE yielded an estimate with average absolute error of $1.46 \times 10^{-2}\pm 9.6 \times 10^{-3}$. For counter-propagating Raman transitions, BRPE provided an estimate with average absolute error of $7.5 \times 10^{-3} \pm 5.3 \times 10^{-3}$. b) Two final posterior distributions produced by BRPE from independent sets of experimental data. The green plot shows a high-confidence, single-peak distribution yielding an accurate estimate with a confidence score of $C_s = 0.99$. In contrast, the red plot's dual-peak distribution results in a low-confidence estimate ($C_s = 0.34$), suggesting a need to rerun BRPE.}
        \label{fig:experiments}
\end{figure*}

\subsection{Experimental Results} \label{sec:experimental_results}

We experimentally implemented Ramsey spectroscopy using BRPE and RPE on a compact cyrogenic trapped ion system \cite{spivey}. The qubit is encoded in the hyperfine ground state of an ${}^{171}\textrm{Yb}^{+}$ ion, with state $\ket{0} \coloneqq {}^2\textrm{S}_{1/2} \ket{F=0, m_F=0}$ and state $\ket{1} \coloneqq {}^2\textrm{S}_{1/2} \ket{F=1, m_F=0}$. Energy splitting between $\ket{0}$ and $\ket{1}$ is $2\pi \times 12.642821$ GHz \cite{olmschenk}. Doppler cooling and resolved sideband cooling \cite{leibfried2} is used to cool the ion to its motional ground state and optical pumping is used for state preparation. The qubit rotations in our Ramsey experiments are implemented using stimulated Raman transitions. We test co-propagating and counter-propagating Raman transitions. In both cases, transitions are driven with $355$ nm picosecond-pulsed laser beams controlled by acousto-optic modulators (AOMs) and steered by micro-electromechanical systems (MEMS) technology \cite{crain}. Co-propagating Raman transitions are implemented with two tones on a single global beam, while counter-propagating transitions are executed using the global beam and one individual-addressing beam.

We first validate the accuracy and precision of BRPE and RPE in our experimental setup and compare average absolute errors. The Raman lasers are configured such that our effective laser frequency, $\omega_L = \omega_0 + \delta$ is artificially offset from $\omega_0 = 2\pi \times 12.642821$ GHz, the transition frequency of our qubit, by an amount $\delta$. An estimate of the offset, $\tilde{\delta}$, and confidence score $0 \leq C_s \leq 1$ is then calculated using BRPE with $M_k=9$. This process is repeated for $-2\pi \times 600 \textrm{Hz} \leq \delta \leq 2\pi \times 600 \textrm{Hz}$ with $2\pi\times50$ Hz increments. 

For both co-propagating Raman transitions and counter-propagating Raman transitions BRPE is highly accurate and precise with $\bar{\varepsilon} =  7.5 \times 10^{-3} \pm 5.3 \times 10^{-3} $ radians for counter-propagating Raman transitions and $\bar{\varepsilon} = 1.46 \times 10^{-2} \pm 9.6 \times 10^{-3}$ radians for co-propagating Raman transitions. These findings are visually represented in Fig.\ref{fig:brpe_experiment}, which charts the BRPE estimates across all tested offset increments. 

Using the same experimental data we calculated average absolute error for RPE to be $8.8 \times 10^{-3} \pm 6.5 \times 10^{-3}$ radians when using counter-propagating Raman transistions and $1.49 \times 10^{-2} \pm 9.9 \times 10^{-3}$ radians when using co-propagating Raman transitions. 

In the regime where $\delta$ is small relative to $\omega_0$, we anticipate the gains from using BRPE to be minimal. Nevertheless, BRPE remains the preferable protocol because, as $\delta$ increases, it provides more accurate and precise estimates as demonstrated in section \ref{sec:simulations}.

Next, we focus on two specific instances of BRPE to illustrate the effectiveness of our confidence metric. Fig. \ref{fig:brpe_confidence} displays two BRPE posterior distributions calculated from independent sets of experimental data. The plot in green represents a typical high-confidence posterior distribution characterized by a distinct, single peak. This strong signal enables BRPE to determine an estimate of $\tilde{\delta} = -2\pi \times 449.23$ Hz with a confidence score of $C_s = 0.99$ for an experimentally introduced offset of $\delta=-2\pi\times450$ Hz. In contrast, the plot in red features a posterior distribution with two peaks of approximately equal amplitude but separated by several hundred Hertz. In this scenario, BRPE is unable to deliver a high-confidence estimate, resulting in $\tilde{\delta}=-2\pi\times245.83$ Hz with a confidence score of $C_s=0.34$ for an experimentally added offset of $\delta=-2\pi\times250$ Hz. This low confidence score is a strong indication to distrust this estimate and rerun BRPE. 

\section{Conclusion}
In this work, we introduced Bayesian Robust Phase Estimation (BRPE), an advanced version of Robust Phase Estimation (RPE). BRPE stands out for its ability to produce estimates with significantly lower average absolute error, requiring fewer experimental trials compared to traditional RPE. Our numerical simulations demonstrated that BRPE can achieve estimates with a $96\%$ lower average absolute error when using a fixed sampling count. Notably, this enhancement is consistent even in conditions involving measurement, depolarization, and C-SPAM errors. In practical applications, we successfully implemented BRPE on a trapped ion quantum computer, confirming its speed and efficiency in line with our simulation predictions.

Looking ahead, our research agenda includes plans to validate BRPE's performance on various hardware platforms, further exploring its resilience against different types of errors. Additionally, we aim to refine our confidence metric, enhancing its reliability. This update will provide experimentalists with a more robust tool to assess the trustworthiness of the estimates, facilitating more accurate and reliable quantum measurements. Finally, in this work we focused on a non-adaptive protocol to ease experimental implementation, but future work will explore how the accuracy and precision of BRPE can be further improved by incorporating adaptive Bayesian techniques.  .

\section*{ACKNOWLEDGEMENTS}
T.H. would like to thank Jacob Whitlow, Jonathan Baker, Abhinav Anand, and Kenneth Rudinger for their helpful discussions and thoughtful suggestions. The work was supported by the Spectator Qubit Army Research Office MURI ( W911NF-18-1-
0218 ), US
Department of Energy ( DE-SC0019294 ), and the National Science Foundation ( OSI-2326810 ).

\bibliography{references}{}
\bibliographystyle{plain}

\vspace{12pt}

\end{document}